\documentclass[twocolumn,preprintnumbers,amsmath,amssymb]{revtex4}  

\usepackage{graphicx}
\usepackage{dcolumn}
\usepackage{bm}
\usepackage{multirow}
\usepackage{color}
\usepackage{amsmath}

\begin{document}
\title{Polarization and time-resolved photoluminescence spectroscopy\\ of excitons in MoSe$_2$ monolayers}

\author{G. Wang$^1$}
\author{E. Palleau$^1$}
\author{T. Amand$^1$}
\author{S. Tongay$^2$}
\author{X. Marie$^1$}
\author{B. Urbaszek$^1$}
\affiliation{%
$^1$ Universit\'e de Toulouse, INSA-CNRS-UPS, LPCNO, 135 Av. de Rangueil, 31077 Toulouse, France}

\affiliation{%
$^2$ School for Engineering of Matter, Transport and Energy, Arizona State University, Tempe, AZ 85287, USA}

\begin{abstract}
We investigate valley exciton dynamics in MoSe$_2$ monolayers in polarization- and time-resolved photoluminescence (PL) spectroscopy at 4K. 
Following circularly polarized laser excitation, we record a low circular polarization degree of the PL of typically $\leq5\%$. 
This is about 10 times lower than the polarization induced under comparable conditions in MoS$_2$ and WSe$_2$ monolayers.  
The evolution of the exciton polarization as a function of excitation laser energy  and power is monitored in PL excitation (PLE) experiments.
Fast PL emission times are recorded for both the neutral exciton of $\leq3$~ps and for the charged exciton (trion) of $12$~ps. 
\end{abstract}


                             
\maketitle
Monolayers (MLs) of the transition metal dichalcogenides (TMDCs) MoS$_2$, MoSe$_2$, WS$_2$ and WSe$_2$ are semiconductors with a direct bandgap in the visible region \cite{Mak:2010a,Splendiani:2010a,Zhang:2014a}. Their optical properties are dominated by excitons, strongly Coulomb-bound electron hole pairs \cite{Tawinan:2012a,Komsa:2012a,Ross:2013a,He:2014a,Ugeda:2014a,Chernikov:2014a,Ye:2014a,Wang:2014a,Klots:2014a}. TMDC MLs have emerged as very promising materials for optical, electronic and quantum manipulation applications \cite{Geim:2013a,Xu:2014a}.  In TMDC  MLs crystal inversion symmetry breaking together with the strong spin-orbit (SO) interaction leads to a coupling of carrier spin and k-space valley physics, i.e., the circular polarization ($\sigma^+$ or $\sigma^-$) of the absorbed or emitted photon can be directly associated with selective carrier excitation in one of the two non-equivalent $K$ valleys ($K^+$ or $K^-$, respectively) \cite{Xiao:2012a, Cao:2012a,Mak:2012a,Sallen:2012a,Kioseoglou:2012a,Jones:2013a,Mak:2014a}. 
\begin{figure}
\includegraphics[width=0.45\textwidth]{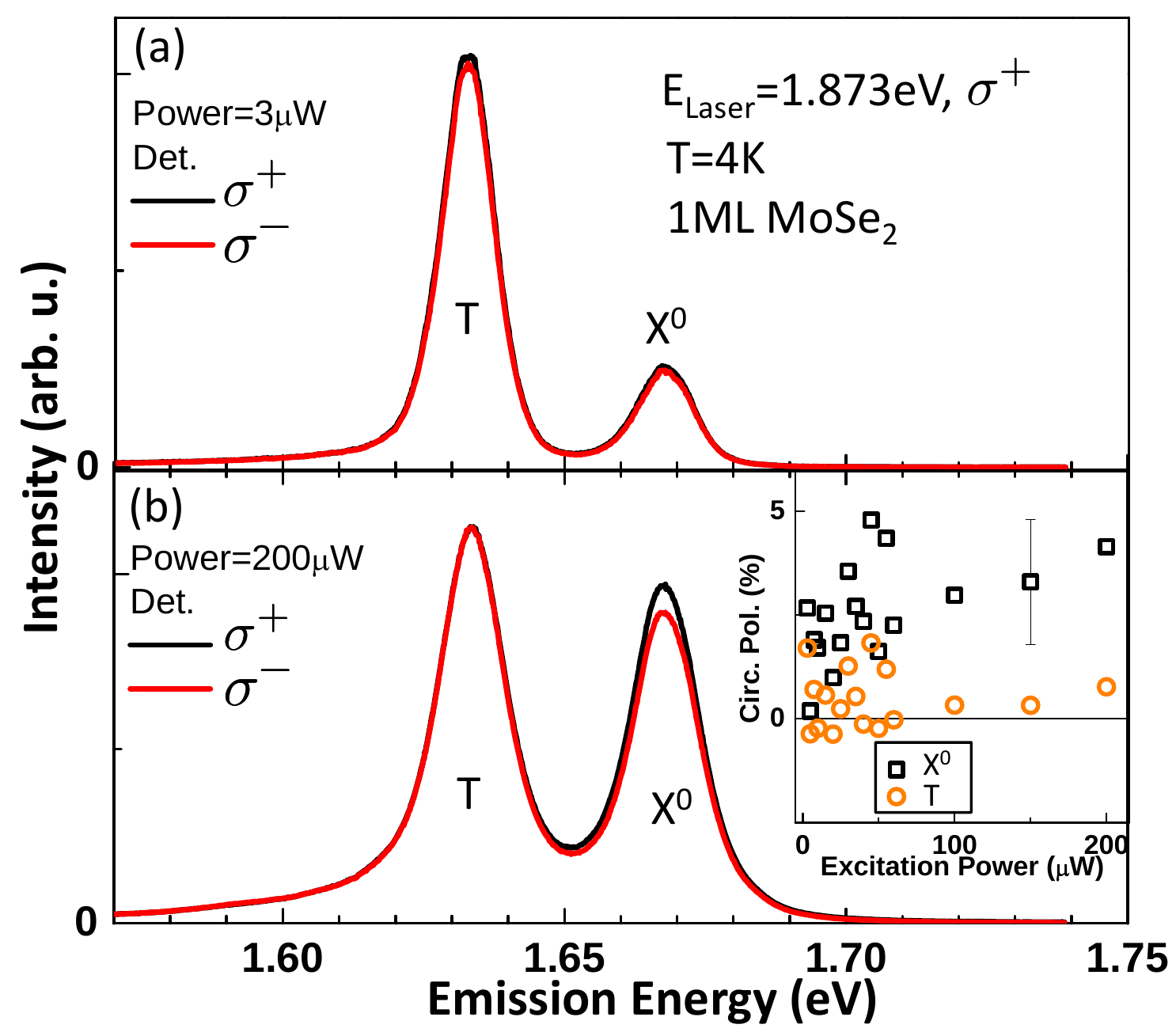}
\caption{\label{fig:fig1} (a) Time-integrated PL intensity of the MoSe$_2$ ML exhibiting both neutral (X$^0$) and charged exciton (T)  optical transitions at T=4 K, $E_{\text{laser}}$~=1.873~eV, laser power 3~$\mu$W, laser $\sigma^+$ polarized, detection $\sigma^+$ (black) and $\sigma^-$ (red). (b) same as (a) but at higher laser power of 200~$\mu$W. Inset: Power dependence of circular polarization degree $P_c$ of PL for X$^0$ (squares) and trions (circles). Typical errors $\pm1.5\%$ are shown.
}
\end{figure}
These chiral optical selection rules leading to strong valley selectivity are expected to be a common feature for MoS$_2$, MoSe$_2$, WS$_2$ and WSe$_2$. High values of the order of $50\%$ for the circular polarization $P_c$ of the stationary photoluminescence (PL) emission corresponding to successful valley polarization have been reported in MoS$_2$  \cite{Mak:2012a,Zeng:2012a,Kioseoglou:2012a,Sallen:2012a}, WSe$_2$ \cite{Jones:2013a,Wang:2014a} and WS$_2$ \cite{Zhu:2014b}, albeit with very different dependences on laser excitation energy i.e. on the excess energy of the initial excitation compared to the exciton emission energy. \\
\indent So ideal conditions for valley polarization generation in PL experiments need to be investigated also for ML MoSe$_2$. This material is very promising for valley index manipulation \cite{Singh:2014a,Kumar:2014a} with bright, well separated emission lines for neutral (X$^0$) and charged excitons (trions T) \cite{Ross:2013a} shown in Fig.~\ref{fig:fig1}, which allows to investigate the valley physics for these complexes individually at low temperature. In time-integrated PL experiments at 4K we record essentially unpolarized emission $P_c\simeq 0$ following excitation as close as 100~meV above the A-exciton with a $\sigma^+$ polarized laser. 
One possible origin for the low PL polarization would be efficient depolarization with a typical time $\tau_s$ much shorter than the PL emission time $\tau$. Our time resolved measurements show PL emission times $\tau$ in the ps range, just as in the case of ML MoS$_2$ \cite{Lagarde:2014a} and WSe$_2$ \cite{Wang:2014b}. This hints at either faster polarization relaxation in MoSe$_2$ in the sub-picosecond range or inefficient optical polarization generation due to anomalies in the bandstructure. \\
\indent MoSe$_2$ ML flakes are obtained by micro-mechanical cleavage of a bulk MoSe$_2$ crystal on 300 nm SiO$_2$ on a Si substrate using viscoelastic stamping \cite{Gomez:2014a}. The ML region is identified by optical contrast and very clearly in PL spectroscopy. Experiments between T=4 and 300~K are carried out in a confocal microscope optimized for polarized PL experiments \cite{Wang:2014b}. The MoSe$_2$ ML is excited by picosecond pulses generated by a tunable frequency-doubled optical parametric oscillator (OPO) synchronously pumped by a mode-locked Ti:Sa laser. The typical pulse and spectral width are 1.6 ps and 3 meV respectively; the repetition rate is 80 MHz. The laser average power is tunable from 2 to 200~$\mu$W. The detection spot diameter is $\approx1\mu$m. For time integrated experiments, the PL emission is dispersed in a spectrometer and detected with a Si-CCD camera. For time-resolved experiments, the PL signal is dispersed by an imaging spectrometer and detected by a synchro-scan Hamamatsu Streak Camera with an overall time resolution of 3 ps. The circular PL polarization $P_c$ is defined as $P_c=(I_{\sigma+}-I_{\sigma-})/(I_{\sigma+}+I_{\sigma-})$, where $I_{\sigma+}(I_{\sigma-})$denotes the intensity of the right ($\sigma^+$) and left ($\sigma^-$) circularly polarized emission. \\
\begin{figure}
\includegraphics[width=0.45\textwidth]{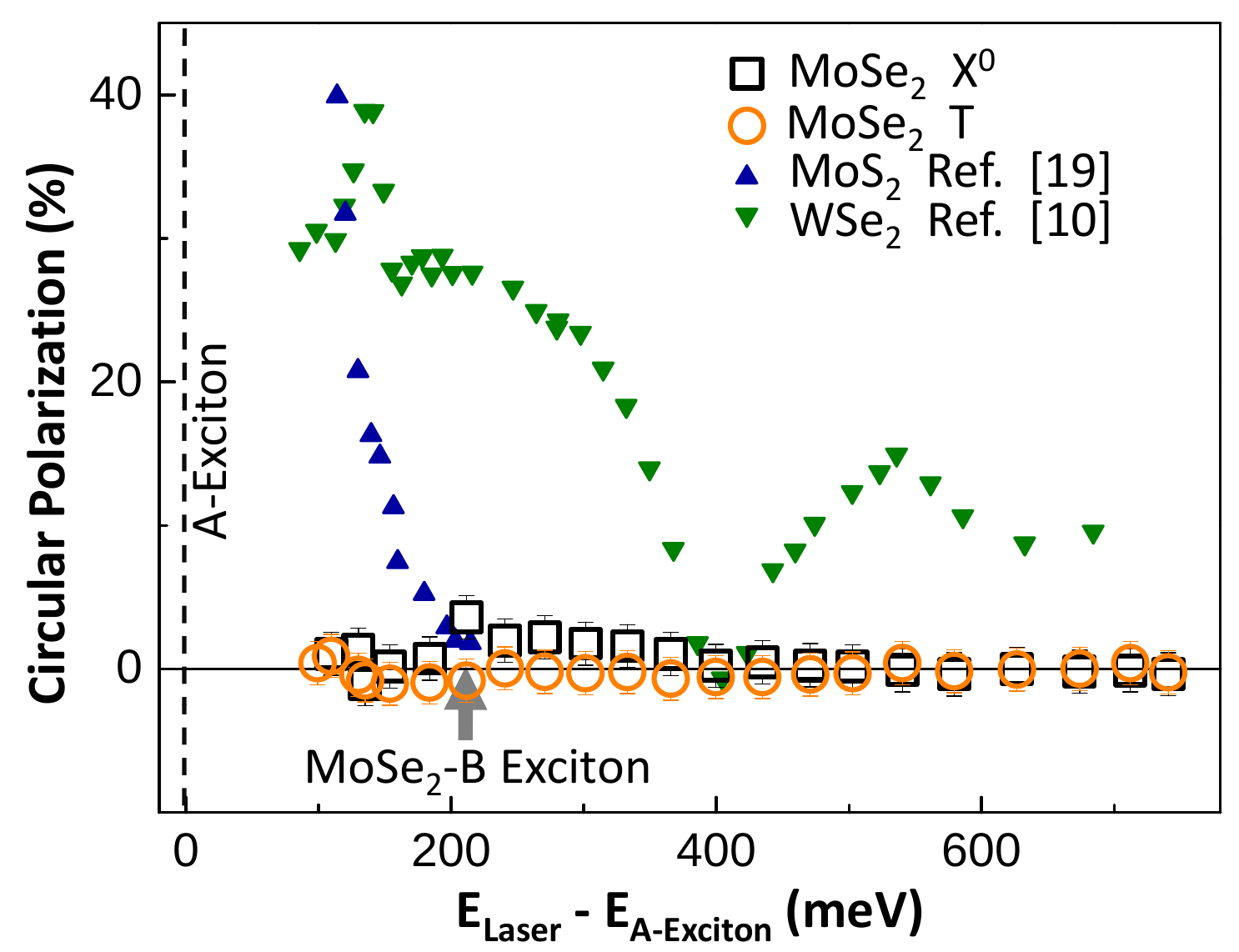}
\caption{\label{fig:fig2} Measured circular polarization $P_c$ of PL for X$^0$ (squares) and trions (circles) as a function the difference between the laser energy and the A-exciton emission in ML MoSe$_2$. For comparison, the results obtained on ML MoS$_2$ (upward triangle) by \textcite{Kioseoglou:2012a} and on ML WSe$_2$ (downward triangle) by \textcite{Wang:2014a} in PLE measurements at 4K for samples on Si/SiO$_2$ are shown, with typical experimental errors of $\pm1.5\%$. As the origin of the energy axis the neutral A-exciton emission is taken for each material, respectively. 
}
\end{figure} 
\indent In order to experimentally verify valley selectivity in interband transitions, we excite the MoSe$_2$ ML with a  circularly $\sigma^+$ polarized laser above the A-exciton energy and we monitor the circular polarization $P_c$ of the resulting PL emission. Here we vary two important  parameters, namely the laser energy and power. In Fig.~\ref{fig:fig1}a we see a typical time-integrated PL spectrum for ML MoSe$_2$ with two prominent peaks. The higher energy peak (FWHM=10~meV) at 1.667~eV has previously been attributed to the neutral exciton X$^0$ \cite{Ross:2013a,Li:2014a,Macneill:2015a}. At 1.633~eV we record the trion emission corresponding to a binding energy of 34~meV due to the strong Coulomb interaction in this material \cite{Ugeda:2014a}. Following $\sigma^+$ polarized laser excitation, the PL emission of both X$^0$ and the trion is equally intense for $\sigma^+$ and $\sigma^-$ polarized emission corresponding to $P_c\simeq0$ within our experimental accuracy for a laser excitation energy of 1.873~eV using low laser power of 3~$\mu$W. \\
\indent The PL polarization in the closely related material MoS$_2$ has been found to depend on laser excitation power \cite{Lagarde:2014a}. Increasing the power of the laser exciting the ML MoSe$_2$ by two orders of magnitude has two main effects, as can be seen in Fig.~\ref{fig:fig1}b: First, the intensity ratio X$^0$:T increase with power, the X$^0$ gains in relative strength. Second, as shown in the inset of Fig.~\ref{fig:fig1}b, we record a very small increase of the circular polarization of X$^0$ up to about $P_c\simeq 4\%$, whereas the trion emission remains unpolarized. This is surprising, as the depolarization mechanisms based on the strong electron-hole Coulomb exchange interaction in these systems suggest a faster valley decay, and hence lower $P_c$ in time integrated measurements, for the X$^0$ than for the trion \cite{Glazov:2014a,Yu:2014a}. \\
\begin{figure}
\includegraphics[width=0.45\textwidth]{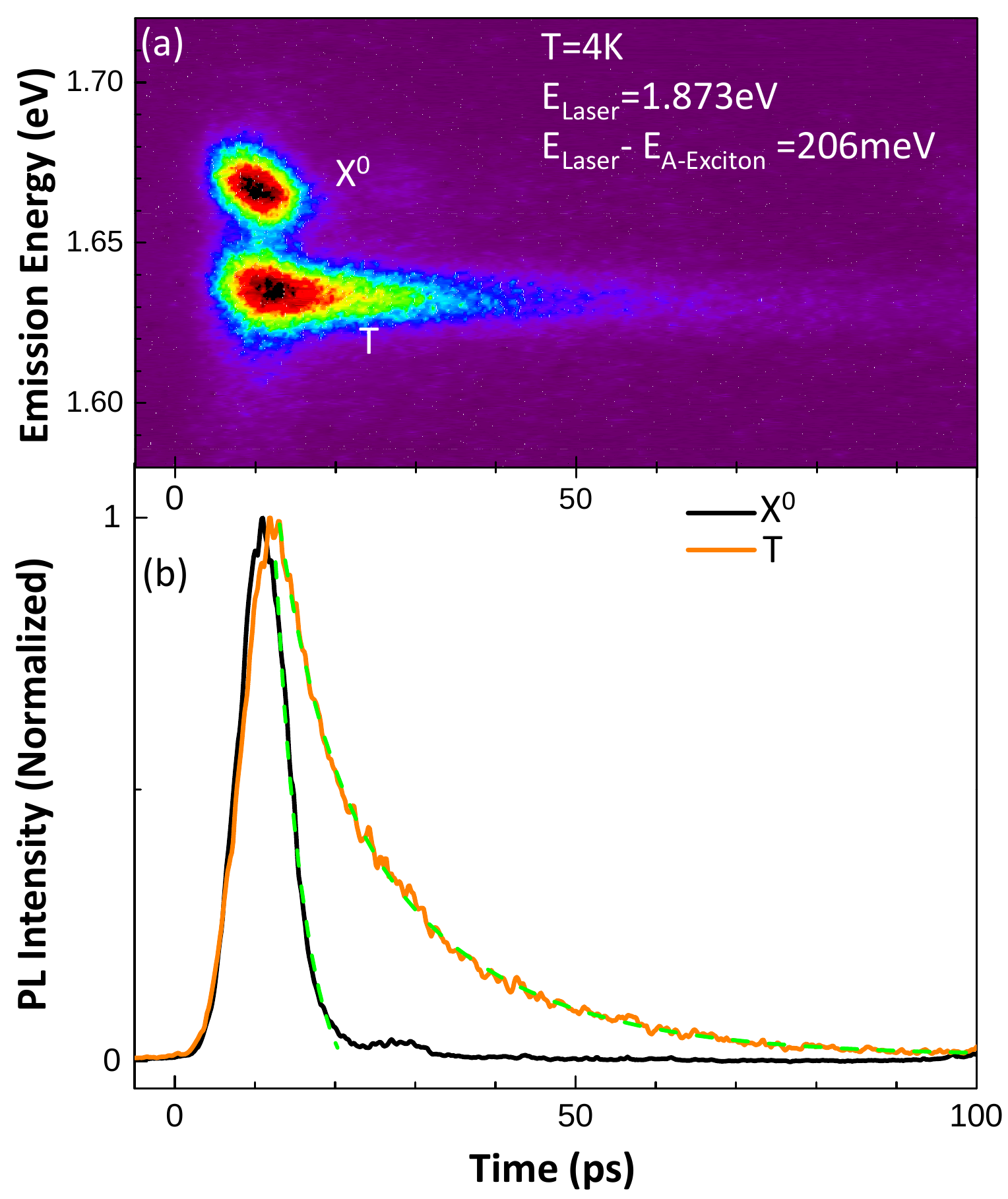}
\caption{\label{fig:fig3} (a) Streak camera image in false color scale of ML MoSe$_2$ PL emission at 4K, the neutral exciton emission (X$^0$) and the trion emission (T) are plotted as a function of time for $E_{\text{laser}} = 1.873~$eV, i.e. $E_{\text{laser}}-E_{\text{A-Exciton}} = 206~$meV. (b) PL dynamics at the central energy of the X$^0$ (black) and T (orange) emission. The dashed lines show mono-exponential fits for the X$^0$ with a single characteristic time of $\tau_{X0}\lesssim3$~ps and for the trion of $\tau_{T}\approx12$~ps.
}
\end{figure} 
\indent The chiral optical selection rules for ML TMDCs are based on the symmetry properties of the K-points at the edge of the Brillouin zone \cite{Xiao:2012a, Cao:2012a}. In most PL experiments the laser excitation is not strictly resonant with the A-exciton, i.e. electron hole pairs (excitons) in energy well above the K-points are generated. 
The exact selection rules away from the K-points are not precisely known as they depend on the particular band-structure of the TMDC. Also strong excitonic effects (binding energies of the order of 0.6~eV \cite{Ugeda:2014a}) have to be taken into account. As a result contributions of k-states of electron-hole pairs far away from the K-points constitute the  wavefunction of the exciton. It is therefore crucial to monitor the circular polarization degree for the PL emission as a function of laser energy. 
Our results are plotted in Fig.~\ref{fig:fig2}: Going from 1.8 to 2.4 eV (i.e. 0.1 to 0.6~eV above the A-exciton) the trion emission remains essentially unpolarized. For the neutral exciton we only see a slight increase in $P_c$ at an energy of 1.9~eV, which corresponds to the excitation of the B-exciton in the same valley. 
These findings shown in Fig.~\ref{fig:fig2} are in stark contrast to the results reported for PLE measurements in MoS$_2$ \cite{Kioseoglou:2012a,Lagarde:2014a} and WSe$_2$ \cite{Wang:2014a}, plotted for comparison in the same figure. For ML MoS$_2$ a circular polarization degree above $40\%$ was reported for laser excitation about 100~meV above the neutral A-exciton \cite{Kioseoglou:2012a}. For ML WSe$_2$ a global maximum for valley polarization and coherence was found 140~meV above the neutral A-exciton transition, which is interpreted as the excitation of the excited exciton level 2s \cite{Wang:2014a}. It is therefore very surprising to see for laser excitation energies as close as 100~meV no measurable valley polarization in MoSe$_2$. We emphasize that similar exciton binding energies of the order of 0.6~eV have been reported for all these monolayers (MoS$_2$, MoSe$_2$, WSe$_2$, WS$_2$ \cite{He:2014a,Ugeda:2014a,Chernikov:2014a,Ye:2014a,Wang:2014a,Klots:2014a} ). Note that experiments at lower laser energy resulted in scattered laser light and pronounced Raman features superimposed with the PL emission and no reliable polarization measurements could be carried out. It is very interesting that the three related materials MoS$_2$, WSe$_2$ and MoSe$_2$ on SiO$_2$/Si substrates show very different dependence of the valley polarization on laser excess energy in PL experiments. No direct link between the valley polarization values and  the amplitude of the SO splitting (140~meV in MoS$_2$, $\approx200~$meV in MoSe$_2$, $\approx400~$meV in WSe$_2$ for the valence band) can be deduced at this stage.\\
\indent The circular polarization degree $P_c$ detected in time integrated measurements depends on the initially generated polarization $P_0$ at t=0, the PL emission time $\tau$ and the polarization decay time $\tau_s$ in a simple model as $P_c(t)=P_0/(1+\tau/\tau_s)$, see for example \cite{Dyakonov:2008a}. With the aim to understand the origin of the unusually low PL polarization of the X$^0$ and the trion at 4K in MoSe$_2$ MLs we perform time resolved PL experiments on exactly the same flake used before for the PLE measurements of Fig.~\ref{fig:fig2}. The results obtained with a Streak camera detection system are shown in Fig.~\ref{fig:fig3}a and b. For the X$^0$ we observe a fast decay of the PL of $\tau \lesssim 3$~ps, essentially within our time resolution. The trion PL emission decays more slowly, as can be clearly seen in Fig.~\ref{fig:fig2}a and b, with a typical decay time of $\tau \simeq 12$~ps . These results are very similar to the fast PL decay reported for the X$^0$ and trion in ML WSe$_2$ \cite{Wang:2014b}, a material for which high $P_c$ values in stationary PL could be achieved. This suggests the following explanations for the absence of valley polarization in ML MoSe$_2$: \\
(i) The initially created polarization $P_0$ is negligible. Here the exact symmetry of the transitions involved in the absorption process needs to be investigated. Excitonic effects play an important role in all TMDC MLs and the small Bohr radius results in a large extension in k-space i.e. the excitonic wavefunction contains contributions from states relatively far away from the K-point, which may not obey the strict, chiral selection rules valid at the K-point.\\
(ii) Although a finite polarization $P_0$ can be generated, the polarization decay time $\tau_s$ is faster than the ps recombination time $\tau$. Using Kerr rotation in ML WSe$_2$, \textcite{Zhucr:2014a} have found $\tau_s$ of the order of 6~ps compared to a PL emission time in the sample of $\tau \lesssim 3$~ps, which still resulted in high PL polarization in cw measurements. Valley depolarization in the ps range was ascribed to valley coupling by strong Coulomb exchange effects \cite{Glazov:2014a,Yu:2014a}. So one possible scenario for ML MoSe$_2$ is $\tau_s \ll\tau \lesssim 3$~ps. ML MoSe$_2$ samples do not show any measurable defect emission at energies below the trion PL, see Fig.~\ref{fig:fig1}, contrary to WSe$_2$ \cite{Wang:2014b} and MoS$_2$ \cite{Sallen:2012a} samples. So the scattering time $\tau^*$ on defects for neutral excitons is expected to be longer. As a result, the polarization lifetime could get shorter as $1/\tau_s=\Omega^2 \tau^*$, where $\Omega$ is the precession frequency about the effective field induced by long-range Coulomb exchange \cite{Glazov:2014a,Maialle:1993a,Dyakonov:2008a}. In this context, the increase of the time-integrated polarization of the neutral exciton as a function of excitation power (inset Fig.~1b) is consistent with a motional narrowing effect. If the exciton density increases, the scattering time $\tau^*$ decreases and as a consequence the spin/valley relaxation time increases inducing a larger polarization. \\
\indent Another strategy to generate valley polarization in MoSe$_2$ is to apply a magnetic field perpendicular to the ML plane and to lift the degeneray between the K$_+$ and K$_-$ valleys. In this configuration up to $P_c=50\%$ have been reported \cite{Macneill:2015a,Li:2014a}, the origin of valley polarization in this case is not attributed to optical valley initialization, but such a large value would be consistent with an ultra-fast (sub-picosecond) valley relaxation time.\\
\indent We thank Iann Gerber and Misha Glazov for fruitful discussions and acknowledge partial funding from ERC Grant No. 306719, ANR MoS2ValleyControl and Programme Investissements d'Avenir ANR-11-IDEX-0002-02, reference ANR-10-LABX-0037- NEXT.

\end{document}